\documentclass[showpacs, oneside, twocolumn, prd, amsmath, amssymb, nofootinbib, superscriptaddress]{revtex4-1}

\usepackage{cases}
\usepackage{amsmath}
\usepackage{amssymb}
\usepackage{amsfonts}
\usepackage{amssymb}
\usepackage{dcolumn}
\usepackage{bm}
\usepackage{bbm}
\usepackage{graphicx}
\usepackage{xcolor}
\usepackage{array}
\usepackage{subfigure}
\usepackage{hyperref}
\usepackage{wasysym}

\newcommand{\be}{\begin{equation}}
\newcommand{\ee}{\end{equation}}
\newcommand{\ba}{\begin{eqnarray}}
\newcommand{\ea}{\end{eqnarray}}

\newcommand{\gsim}{\mathrel{\hbox{\rlap{\lower.55ex \hbox {$\sim$}}
			\kern-.3em \raise.4ex \hbox{$>$}}}}
\newcommand{\lsim}{\mathrel{\hbox{\rlap{\lower.55ex \hbox {$\sim$}}
			\kern-.3em \raise.4ex \hbox{$<$}}}}

\hypersetup{colorlinks=true,
	breaklinks=true,
	pdfstartview=Fit,
	linkcolor=blue,
	citecolor=blue,
	urlcolor=blue}

\begin{document}
\title{When Primordial Black Holes {from Sound Speed Resonance} Meet a Stochastic Background of Gravitational Waves}

\author{Yi-Fu Cai}
\email{yifucai@ustc.edu.cn}
\affiliation{Department of Astronomy, School of Physical Sciences, University of Science and Technology of China, Hefei, Anhui 230026, China}
\affiliation{CAS Key Laboratory for Researches in Galaxies and Cosmology, University of Science and Technology of China, Hefei, Anhui 230026, China}
\affiliation{School of Astronomy and Space Science, University of Science and Technology of China, Hefei, Anhui 230026, China}

\author{Chao Chen}
\email{cchao012@mail.ustc.edu.cn}
\affiliation{Department of Astronomy, School of Physical Sciences, University of Science and Technology of China, Hefei, Anhui 230026, China}
\affiliation{CAS Key Laboratory for Researches in Galaxies and Cosmology, University of Science and Technology of China, Hefei, Anhui 230026, China}
\affiliation{School of Astronomy and Space Science, University of Science and Technology of China, Hefei, Anhui 230026, China}

\author{Xi Tong}
\email{xtongac@connect.ust.hk}
\affiliation{Department of Physics, The Hong Kong University of Science and Technology, Clear Water Bay, Kowloon, Hong Kong, China}

\author{Dong-Gang Wang}
\email{wdgang@strw.leidenuniv.nl}
\affiliation{Lorentz Institute for Theoretical Physics, Leiden University, 2333 CA Leiden, The Netherlands}
\affiliation{Leiden Observatory, Leiden University, 2300 RA Leiden, The Netherlands}

\author{Sheng-Feng Yan}
\email{sfyan22@mail.ustc.edu.cn}
\affiliation{Department of Astronomy, School of Physical Sciences, University of Science and Technology of China, Hefei, Anhui 230026, China}
\affiliation{CAS Key Laboratory for Researches in Galaxies and Cosmology, University of Science and Technology of China, Hefei, Anhui 230026, China}
\affiliation{School of Astronomy and Space Science, University of Science and Technology of China, Hefei, Anhui 230026, China}

\begin{abstract}
	As potential candidates of dark matter, primordial black holes (PBHs) are within the core scopes of various astronomical observations.
	In light of the explosive development of gravitational wave (GW) and radio astronomy, we thoroughly analyze a stochastic background of cosmological GWs, induced by overly large primordial density perturbations, with several spikes that was inspired by the sound speed resonance effect and can predict a particular pattern on the mass spectrum of PBHs.
	With a specific mechanism for PBHs formation, we for the first time perform the study of such induced GWs that originate from {\it both the inflationary era and the radiation-dominated phase}.
	We report that, besides the traditional process of generating GWs during the radiation-dominated phase, the contribution of the induced GWs in the sub-Hubble regime during inflation can become significant at the critical frequency band because of a narrow resonance effect.
	All contributions sum together to yield a specific profile of the energy spectrum of GWs that can be of observable interest in forthcoming astronomical experiments.
	Our study shed light on the possible joint probe of PBHs via various observational windows of multimessenger astronomy, including the search for electromagnetic effects with astronomical telescopes and the stochastic background of relic GWs with GW instruments.
\end{abstract}

\pacs{98.80.Cq, 11.25.Tq, 74.20.-z, 04.50.Gh}

\maketitle

\section{Introduction}

Primordial black holes (PBHs) are hypothetical objects predicted by many fundamental theories to form soon after the big bang, and hence the search for PBHs offers an inspiring possibility to probe physics in the early Universe \cite{Zeldovich:1966, Hawking:1971ei, Carr:1974nx}. Since they could be a possible candidate for dark matter (DM), the studies on cosmological implications of PBHs are crucial for cosmologists \cite{Ivanov:1994pa, Carr:2016drx, Gaggero:2016dpq, Inomata:2017okj, Georg:2017mqk, Fuller:2017uyd, Kovetz:2017rvv, Cai:2018tuh, Nakama:2018utx, Ballesteros:2018wlw, Carr:2018poi, Deng:2018wmy, Dalianis:2018ymb}. Recently, it was pointed out that the PBHs might be responsible for some gravitational wave (GW) events \cite{Bird:2016dcv, Clesse:2016vqa, Sasaki:2016jop, Nakamura:1997sm}, which were detected by GW instruments such as the LIGO \cite{Abbott:2016blz}. This observational possibility motivates many theoretical mechanisms generating PBHs, which often require a power spectrum of primordial density perturbations to be suitably large on certain scales that are associated with a particularly tuned background dynamics of the quantum fields in the early Universe (e.g. see \cite{GarciaBellido:1996qt, Garcia-Bellido:2017mdw, Domcke:2017fix, Kannike:2017bxn, Carr:2017edp, Ballesteros:2017fsr, Hertzberg:2017dkh, Franciolini:2018vbk, Kohri:2018qtx, Ozsoy:2018flq, Biagetti:2018pjj} for studies within inflation, see \cite{Chen:2016kjx, Quintin:2016qro} for discussions within bounce, and see \cite{Khlopov:2008qy, Sasaki:2018dmp} for recent comprehensive reviews).

Primordial density perturbations, which seeded the large-scale structure (LSS) of the Universe, are usually thought to arise from quantum fluctuations during a dramatic phase of expansion at early times, as described by inflationary cosmology, from which  a nearly scale-invariant power spectrum with a standard dispersion relation is obtained \cite{Mukhanov:1990me}. This was confirmed by various cosmological measurements such as the cosmic microwave background (CMB) radiation \cite{Ade:2015lrj, Akrami:2018odb} and LSS surveys at extremely high precision. It is interesting to note that, although density and tensor perturbations evolve independently through the early Universe at linear level, they couple with each other nonlinearly and hence can induce either the non-Gaussianities or stochastic background of relic GWs \cite{Ananda:2006af, Baumann:2007zm}. If these density perturbations could form PBHs, then it becomes possible to constrain primordial non-Gaussianities with PBHs \cite{Bullock:1996at, Ivanov:1997ia, Saito:2008em, Seery:2006wk, Byrnes:2012yx, Young:2013oia}. Moreover, it is intriguing to search for PBHs via the measurements of the stochastic GW background induced by overly large primordial density perturbations \cite{Saito:2008jc, Saito:2009jt, Bugaev:2010bb, Garcia-Bellido:2016dkw, Inomata:2016rbd, Garcia-Bellido:2017aan, Kohri:2018awv, Bartolo:2018evs, Cai:2018dig, Bartolo:2018rku, Wang:2018yql, Unal:2018yaa, Inomata:2018epa, Clesse:2018ogk}. The connection between PBHs and the induced GWs has been extensively studied in the literature, which mainly focused on the GW generation during the radiation-dominated phase when PBHs were formed. So far, the GW generation during the inflationary phase remains unclear, which is one of the main subjects of the present study.


Recently, a novel mechanism for PBH formation, called sound speed resonance (SSR), was proposed in \cite{Cai:2018tuh}, where an oscillating sound speed square leads to the nonperturbative parametric amplification of certain perturbation modes during inflation.
As a result, the power spectrum of the primordial density perturbations has a narrow major peak on small scales, while it remains nearly scale invariant on large scales as predicted by inflationary cosmology. Note that several minor peaks of the power spectrum of the primordial density perturbations on smaller scales are also predicted by this novel mechanism.
In \cite{Cai:2018tuh} it was found that the formation of PBHs caused by the resulting peaks in SSR can be very efficient, which could be testable in future observational experiments.

Moreover, the enhanced primordial density perturbations are expected to induce large GWs signals according to the second-order cosmological perturbation theory.
Motivated by the aforementioned reasons, in the present paper we turn to study the second-order GWs caused by the primordial density perturbations with peaks in the SSR scenario.
Making use of this new mechanism, we perform a full analysis of the GW signal that evolves through the inflationary era until the present Universe.

First, we analyze the GWs induced by the enhanced primordial density perturbations when the relevant modes reenter the Hubble horizon during the radiation-dominated era.
Afterwards, we study GWs induced by the modes of primordial density perturbations at the super-Hubble scales during the inflationary era, which is often omitted in other works. Our calculation reveals that this part of the contribution is usually suppressed by the slow-roll parameter when compared with that of the radiation-dominated era, but it might be important if the slow-roll condition could be violated for a short while during inflation. Finally, we compute the GWs induced by quantum fluctuations that remain inside the Hubble horizon during inflation, and we find that this sub-Hubble contribution can become very significant at the critical frequency band due to the narrow resonance effect.
We interestingly observe that, although the spectrum is damping out at small scales due to a blue tilt, there exists a narrow window where the induced GWs can be resonantly enhanced, and the resulting amplitude can be sizable when compared with the one derived in the radiation-dominated phase.
Therefore, this work provides for the first time a comprehensive study on the stochastic GWs background induced during both the inflationary and the radiation-dominated eras.
The resulting signal provides a new target for various future GW experiments, which also serves as an independent window for probing PBHs in the Universe.


The article is organized as follows. In Sec. \ref{sec:SSRspikes}, we put forward a novel parametrization form of the power spectrum of primordial density perturbations that allows for several spikes and discuss the associated realization from the perspective of the SSR mechanism. In Sec. \ref{sec:inducedGW} we work out the GWs induced by primordial density perturbations during the radiation-dominated phase, and by the super-Hubble modes and the sub-Hubble modes during inflation, respectively. In Sec. \ref{sec:OmegaGW}, we derive the energy spectra of the induced stochastic GWs associated with PBH formation and perform a comparison with the observational ability of present and forthcoming GW experiments.
Finally, we conclude with a discussion in Sec. \ref{sec:conclusion}.
The detailed calculation of induced GWs is presented in the Appendixes.
Throughout the article, we adopt the natural units $c=\hbar=1$ and the reduced Planck mass is defined as $M_p^{-2}=8\pi G$.


\section{Sound Speed Resonance and Power Spectrum with Peaks} \label{sec:SSRspikes}

To generate PBHs within inflationary cosmology, the key point is to amplify the amplitude of primordial density perturbations for certain ranges of modes. For most of the theoretical mechanisms studied in the literature, it requires a manifest enhancement around a unique comoving wave number, and accordingly, the mass spectrum of PBHs displays a single peak around a critical mass scale. However, it was recently pointed out in \cite{Carr:2018poi} that the PBHs are likely to have an extended mass spectrum, in particular with multiple peaks, which has crucial implications for the interpretation of the observational constraints. This novel phenomenon happens to be realized also in \cite{Cai:2018tuh} in terms of the SSR mechanism. We note that, for the sound speed, the first peak of the resonant power spectrum, which corresponds to the lowest comoving wave number, makes the larger contribution to the PBH mass spectrum, but the rest would affect the whole profile, in particular the tail of the mass spectrum.

The causal mechanism of generating the power spectrum suggests that primordial density perturbations initially emerge inside the Hubble radius, and then exit in the primordial epoch, and eventually reenter at late times. One often uses a gauge-invariant variable $\zeta$, the curvature perturbation in the comoving gauge, to depict the primordial inhomogeneities. For convenience, one can introduce a canonical variable $v \equiv z\zeta$, where $z\equiv \sqrt{2\epsilon}a/c_s$ \cite{ArmendarizPicon:1999rj, Garriga:1999vw}. Note that $\epsilon \equiv -\dot H/H^2$ is often regarded as the Hubble slow-roll parameter, $H\equiv \dot{a}/a$ is the Hubble parameter, and $c_s$ is the sound speed parameter of the primordial Universe. The evolution of one Fourier mode for this variable $v_k(\tau)$ satisfies
$ v_k'' + \big( c_s^2 k^2 - z''/z \big) v_k=0 $,
where the prime represents for the derivative with respect to the conformal time $\tau$.

To produce PBHs within inflationary cosmology, it requires a dramatic amplification of the primordial curvature perturbations for certain scales. In \cite{Cai:2018tuh}, a novel mechanism was proposed by introducing an oscillating correction to the sound speed parameter. In particular a parametric amplification of curvature perturbations caused by resonance with oscillations in the sound speed parameter, which provides an efficient way to enhance the primordial power spectrum around the astrophysical scales where PBHs could account for DM in the current experimental bounds. Such an oscillation correction could arise when inflation models are embodied in UV-complete theories, such as D-brane dynamics in string theory \cite{Silverstein:2003hf, Alishahiha:2004eh}, or by integrating out heavy modes from the effective field theory viewpoint \cite{Achucarro:2010da, Achucarro:2012sm}, or purely from a phenomenological construction \cite{Cai:2009hc, Cai:2009hw, Cai:2008if}.

Specifically, the sound speed parameter can be parametrized as
$c_s^2 = 1 -2 \xi [ 1-\cos(2k_*\tau) ]$ with $\tau>\tau_i$,
where $\xi$ is a small dimensionless quantity that measures the oscillation amplitude and $k_*$ is the oscillation frequency. Note that $\xi < 1/4$ is required such that $c_s$ is positively definite, and the oscillation begins at $\tau_i$, where $k_*$ needs to be deep inside the Hubble radius with $|k_*\tau_i|\gg1$. Moreover, we set $c_s=1$ before $\tau_i$ and assume that it can transit to oscillation smoothly for simplicity.
In this mechanism, the perturbation equation can be reexpressed in the form of a Mathieu equation \cite{Cai:2018tuh}, which then gives rise to a narrow parametric resonance; i.e., the perturbation modes in the neighborhood of the characteristic scale $k_*$ can be exponentially enhanced, while the power spectrum on large scales remains nearly scale invariant as predicted by the standard inflationary cosmology. However, it is interesting to observe that this mechanism also predicts several secondary peaks on the scales with integer times of $k_*$. Accordingly, by taking these peaks in the power spectrum into account, one can parametrize its form as follows:
\begin{align} \label{Pzeta}
P_\zeta(k) & = A_s \left( \frac{k}{k_p} \right)^{n_s - 1} \bigg\{1 + \frac{\xi k_*}{2} e^{- \xi k_* \tau_i} \nonumber \\
& \times \Big[ \delta(k - k_*) + \sum_{n=2}^{\infty} a_n \delta(k - n k_*) \Big] \bigg\} ~,
\end{align}
where $A_s = H^2/8 \pi^2 \epsilon M_p^2 $ is the amplitude of the power spectrum predicted by the conventional inflationary paradigm, $\epsilon$ is the Hubble slow-roll parameter as mentioned above, and $n_s$ is the spectral index at the pivot scale $k_p = 0.05 {\rm Mpc}^{-1} $ \cite{Akrami:2018odb}. The resonant enhancements are characterized by the delta functions inside the square brackets in the second line of Eq.~\eqref{Pzeta}, where the amplitude of $n$th peak relative to the first one is quantified by a dimensionless parameter $a_n$. Since we work in the perturbative regime, the height of the peaks in $ P_\zeta(k) $ should be no more than unity. This condition constrains the upper bounds for the amplitudes of the induced GWs generating from the inflationary era and the radiation-dominated era.

To illustrate the technique of calculating GWs induced by the process of PBH production with multiple spikes, throughout the whole analysis we will only take into account the second and third peaks that are located at $2 k_*$ and $3 k_*$ respectively. This is because the amplitudes of other peaks at higher order are exponentially suppressed by a factor of about $O(10^{-8})$ in the SSR mechanism and then soon be out of observable interest. According to the cosmological perturbation theory, different $k$-modes of linear density perturbations can be mixed with each other nonlinearly and this mixing can play the role of generating tensor perturbations at second order \cite{Ananda:2006af, Baumann:2007zm}. 

In the literature, there were extensive studies on the GWs induced by a single-peak pattern of the power spectrum of primordial density perturbations with the scalar modes reentering the Hubble radius during the postinflationary phase \cite{Saito:2008jc, Saito:2009jt, Bugaev:2010bb, Garcia-Bellido:2016dkw, Inomata:2016rbd, Garcia-Bellido:2017aan, Kohri:2018awv, Bartolo:2018evs, Cai:2018dig, Bartolo:2018rku, Wang:2018yql, Unal:2018yaa, Inomata:2018epa, Clesse:2018ogk}. 
For some specific models, the induced GWs associated with PBHs can also be generated during inflation \cite{Cheng:2016qzb, Cheng:2018yyr}.
These works have successfully demonstrated that a possible measurement of the stochastic background of GWs in future astronomical experiments could provide a powerful window to search for PBHs or to constrain the parameter space. However, it is crucial to notice that, this observational attempt requires a precise quantification of the profiles of energy spectra of the induced GWs, which takes into account of the enhancement effects from both inflation and later phases.

To address the aforementioned issues, in the present article we perform a much more comprehensive analysis on how the GWs were induced by a resonantly enhanced power spectrum that began from the sub-Hubble regime of the inflationary era, then evolved to the super-Hubble regime during inflation, and finally re-entered the Hubble horizon during the radiation-dominated phase. The PBHs form in the radiation-dominated phase, but the GWs can be continuously produced throughout the whole evolution as shall be seen in the following section.

\section{Gravitational waves generated in a primordial Universe} \label{sec:inducedGW}

In this section, we present a complete analysis of the induced GWs throughout the whole evolution of a primordial Universe. In order to make a comparison with the pioneering works in the literature, we would like to show the results backward in time. We begin by discussing the induced GWs when the scalar modes reenter the Hubble horizon during the radiation-dominated era.

\subsection{Radiation-dominated era}

In the Newtonian gauge, the line element in the perturbed metric is expressed as \cite{Ananda:2006af, Inomata:2016rbd, Kohri:2018awv}
\be
d^2 s = a^2(\tau) \left\{ -(1 - 2 \Phi) d\tau^2 + \left[ (1 + 2 \Phi) \delta_{ij} + \frac12 h_{ij} \right] dx^i dx^j \right \} ~,
\ee
where $\tau$ is the conformal time, $\Phi$ is the Bardeen potential, and $h_{ij}$ is the tensor perturbation up to second order.
The equation of motion (EoM) for the induced GWs in the Fourier space can be written as follows:
\be \label{EOM_hk}
h^{\lambda''}_{\mathbf{k}}(\tau) + 2 \mathcal{H} h^{\lambda'}_{\mathbf{k}}(\tau) + k^2 h^\lambda_{\mathbf{k}}(\tau) = S^\lambda_{\mathbf{k}}(\tau) ~,
\ee
where $\mathcal{H} \equiv a'/a$ is the comoving Hubble parameter; $\lambda = +, \times$ denote two polarization modes of the GWs; and the source term $S^\lambda_{\mathbf{k}}(\tau) $ during the radiation-dominated era is given by \cite{Kohri:2018awv, Baumann:2007zm, Ananda:2006af, Bartolo:2018rku}
\begin{align} \label{Sk_RD}
S^\lambda_{\mathbf{k}}(\tau) &= 4 \int \frac{\mathrm{d}^3\mathbf{p}}{(2\pi)^3} \mathbf{e}^\lambda(\mathbf{k},\mathbf{p})
\Big[ 3 \Phi_\mathbf{p} \Phi_{\mathbf{k} - \mathbf{p}} + \mathcal{H}^{-2} \Phi_\mathbf{p}' \Phi_{\mathbf{k} - \mathbf{p}}' \nonumber\\
& + \mathcal{H}^{-1} \Phi_\mathbf{p}' \Phi_{\mathbf{k} - \mathbf{p}} + \mathcal{H}^{-1} \Phi_\mathbf{p} \Phi_{\mathbf{k} - \mathbf{p}}' \Big] ~,
\end{align}
where the projections are defined as $\mathbf{e}^\lambda(\mathbf{k},\mathbf{p}) \equiv e^\lambda_{lm}(\mathbf{k}) p_l p_m$ \cite{Saito:2009jt}. They take the form of $(1/\sqrt{2}) p^2 (1 - \mu^2) \cos 2\varphi$ for $\lambda= + $ and $ (1/\sqrt{2}) p^2 (1 - \mu^2) \sin 2\varphi $ for $ \lambda=\times $, where $ \mu \equiv \mathbf{k} \cdot \mathbf{p} / |\mathbf{k}| \cdot |\mathbf{p}|= \cos\theta $, $\theta$ is the angle between the wave vector $ \mathbf{k} $ (of the induced GWs) and $ \mathbf{p} $ (of the source), and $ \varphi $ is the azimuth angle of $ \mathbf{p} $.
The basic formulas above are presented in Appendix \ref{Basic_APP} in more detail.
Accordingly, the EoMs for the induced GWs \eqref{EOM_hk} are actually not equivalent for different polarizations, and hence we keep the indices of polarizations in the expressions.
During the radiation-dominated era, the Bardeen potential $\Phi_\mathbf{p} $ in the source term \eqref{Sk_RD} is related to the primordial comoving curvature perturbation $\zeta_\mathbf{p}$ via the relation $\Phi_\mathbf{p} = (2/3) T(k\tau) \zeta_\mathbf{p}$, and the transfer function $T(k\tau) $ can be solved by the EoM for the Bardeen potential (see Appendix \ref{PS_RD_APP} for details) \cite{Ananda:2006af, Baumann:2007zm}.

The special solution of the induced GWs \eqref{EOM_hk} can be determined by the Green function method.
Note that, while the induced GWs generated during the radiation-dominated phase should also depend on the initial conditions that were inherited from inflationary phase, the main contribution comes from the source term. 
From the EoM \eqref{EOM_hk} and the source term \eqref{Sk_RD}, one can see that the two-point correlation function of the induced GWs can be roughly estimated as the square of the two-point correlation function of $ \zeta_\mathbf{p} $ \cite{Baumann:2007zm, Ananda:2006af}.
After some lengthy calculations, one can derive the power spectrum of the induced GWs during the radiation-dominated phase as follows \cite{Kohri:2018awv, Baumann:2007zm, Ananda:2006af, Bartolo:2018rku}:
\begin{align}\label{Ph_PzetaPzeta_RD}
P^{\text{RD}}_h(k, \tau) &= \int^\infty_0 \mathrm{d}y \int^{1 + y}_{|1 - y|} \mathrm{d}x \left[\frac{4 y^2 - (1 + y^2 - x^2)^2}{4 x y} \right]^2 \nonumber\\
& \times P_\zeta(k x) P_\zeta(k y) F_{\text{RD}}(k,\tau,x,y) ~,
\end{align}
where we have introduced the variables $x = |\mathbf{k} - \mathbf{p}| / k$ and $y = p/k $. Moreover, the function $ F_{\text{RD}}(k,\tau,x,y) $ is given by
\begin{align}
&F_{\text{RD}}(k,\tau,x,y) \nonumber
\\
&= \frac {4} {81} \frac{1}{z^2} \big[ \cos^2(z)~\mathcal{I}_c^2 + \sin^2(z)~\mathcal{I}_s^2  + \sin( 2 z ) \mathcal{I}_c \mathcal{I}_s \big] ~,
\end{align}
where $ z = k \tau $ has been introduced.
The expressions for the functions $ \mathcal{I}_c(x,y) $ and $ \mathcal{I}_s(x,y) $ are provided in the Appendix \ref{PS_RD_APP} for detailed information.
The formalism \eqref{Ph_PzetaPzeta_RD} is the general expression to calculate the power spectrum of the GWs induced by the primordial curvature perturbation $\zeta$ when the associated Fourier modes of $ \zeta $ reenter the Hubble horizon during the radiation-dominated era. The same modes of curvature perturbations can also induce overly large density fluctuations that eventually could form PBHs after reentering the Hubble horizon during the radiation-dominated phase.

We comment that a significant property of the GWs induced by the multispike power spectrum of the primordial curvature perturbations is that the corresponding wave band is wider than that of the single-peak case \cite{Cai:2019amo}.
This can lead to differences in the parametrization of the energy spectrum for the induced GWs, which shall be discussed in detail in Sec. \ref{sec:OmegaGW}. Note that the average amplitude of the induced GWs is not altered by the multispike case, as this amplitude mainly relies on the first peak that in the SSR mechanism is always the dominant one.

\subsection{Inflationary era}

In this part, we investigate the induced GWs during the inflationary era, which is often omitted in previous works. In the SSR mechanism, the curvature perturbations are amplified deep inside the Hubble horizon at the beginning time $\tau_i$ while the sound speed starts oscillating. Then, the curvature perturbations are exponentially enhanced due to the narrow resonance effect until the relevant modes exit the horizon, and they are frozen at the superhorizon scales. In principle, the GWs could be induced in both the sub-Hubble regime and the super-Hubble regime. Thus, the total contributions to the induced GWs during the inflationary era consist of these two regimes. First of all, we calculate the power spectrum for the GWs induced by the super-Hubble modes of the primordial curvature perturbations with the multiple peaks.
Afterwards, we also calculate the power spectrum of GWs induced by the sub-Hubble modes of the primordial curvature perturbations which remain the quantum nature.

\subsubsection{Super-Hubble modes}

During the inflationary era, the perturbed inflaton can provide the anisotropic stress which then sources the GWs nonlinearly.  In this case, the source term in \eqref{EOM_hk} becomes
\be \label{Source_Inf}
S^\lambda_{\mathbf{k}}(\tau)
=
4 \frac{c_s^2}{M_p^2}
\int \frac{\mathrm{d}^3 \mathbf{p}}{(2 \pi)^{3}} \mathbf{e}^\lambda(\mathbf{k},\mathbf{p})
\delta\phi_{\mathbf{p}}(\tau) \delta\phi_{\mathbf{k} - \mathbf{p}}(\tau) ~,
\ee
where $\delta\phi_{\mathbf{p}}(\tau)$ is the perturbed inflaton in the Fourier space. We work in the spatially flat gauge, where the field fluctuation $\delta\phi$ is related to the curvature perturbation $\zeta$ via $\zeta = H\delta\phi / \dot{\phi} = \delta\phi / \sqrt{2 \epsilon} M_p$ at the super-Hubble scales. After the lengthy calculations similar to \eqref{Ph_PzetaPzeta_RD}, we obtain the power spectrum for the induced GWs as follows:
\begin{align} \label{Ph_PzetaPzeta_Inf}
P^{\text{Super}}_h(k, \tau)
&=
\int^\infty_0 \mathrm{d}y \int^{1 + y}_{|1 - y|} \mathrm{d}x \left[\frac{4 y^2 - (1 + y^2 - x^2)^2}{4 x y} \right]^2 \nonumber
\\& \times
P_\zeta(k x) P_\zeta(k y) F_{\text{Inf}}(k,\tau,u) ~.
\end{align}
The function  $ F_{\text{Inf}}(k,\tau,u) $ here is
\begin{align} \label{Finf}
F_{\text{Inf}}(k,\tau,u)
&\simeq
16 \epsilon^2 u^{-2} \nonumber
\\& \times
\Big[ u + z \cos(z + u) - \sin(z + u) \Big]^2 ~,
\end{align}
where $ u \equiv k/p_*$, and $ p_* $ is the characteristic frequency in the SSR mechanism. Comparing the formula \eqref{Ph_PzetaPzeta_Inf} to \eqref{Ph_PzetaPzeta_RD}, we can see that the difference between two power spectra of the induced GWs is from the functions $ F_{\text{Inf}}(k,\tau,u) $ and $ F_{\text{RD}}(k,\tau,x,y) $, since the source terms during the inflationary era \eqref{Source_Inf} and radiation-dominated era \eqref{Sk_RD} are different. We notice that the function $ F_{\text{Inf}}(k,\tau,u) $ involves the slow-roll parameter $ \epsilon $, and hence, the magnitude of the power spectrum \eqref{Ph_PzetaPzeta_Inf} is sensitive to the value of $ \epsilon $. Consequently, the energy spectrum of the GWs induced by the super-Hubble modes is about $ \epsilon^2 $ suppressed comparing with the GWs induced during the radiation-dominated era; see Sec. \ref{sec:OmegaGW}. 
The explicit expression for $P^{\text{Super}}_h(k,\tau) $ induced by the single-peak pattern and multipeak pattern of $P_\zeta(k)$ can be found in \eqref{Ph_Single_APP} and \eqref{Ph_Spikes_APP} of Appendix \ref{PS_Infl_APP}. According to our analysis, the power spectrum for the induced GWs in the super-Hubble regime is frozen at the end of inflation, and $P^{\text{Super}}_h(k,\tau_{\text{end}})$ is shown in Fig. \ref{Ph_Inflation_Fig}, where $\tau_{\text{end}}$ is the conformal time at the end of inflation, i.e., $\tau_{\text{end}}\rightarrow0$. It is easy to see that the peak of the power spectrum \eqref{Ph_PzetaPzeta_Inf} is located at $k\simeq p_*$, where the amplitude arrives at $10^{-12}$, and we use the values of parameters as $\epsilon = 10^{-3}$, $v\equiv-p_*\tau_i=200$, $ \xi = 0.1$, and $n_s=0.968$, which are in agreement with the latest CMB observations \cite{Akrami:2018odb}. However, the effects of multiple peaks are no longer manifest in the tail of the power spectrum due to the suppression of the higher peaks by a factor of about $O(10^{-8})$.

\subsubsection{Sub-Hubble modes}

In this part, we continue to study the induced GWs from the quantum fluctuations of the perturbed inflaton $ \delta\hat{\phi}_\mathbf{p} $ inside the Hubble horizon during the inflationary era. Due to the narrow resonance effect in the SSR mechanism, the sub-Hubble modes of $ \delta\hat{\phi}_\mathbf{p} $ in the neighborhood of the characteristic scale $p_*$ can be exponentially amplified, leading to the first major peak in the curvature power spectrum $P_\zeta(k)$ \eqref{Pzeta}. Since the sub-Hubble modes are time dependent, it is not easy to calculate the four-point correlator of the $ \delta\hat{\phi}_\mathbf{p} $ as the previous calculations. Moreover, the solution of the modes in the neighborhood of the characteristic scale $p_*$ is a complicated combination of Mathieu functions \cite{Cai:2018tuh}. Therefore, we use a semi-analytical method to compute the power spectrum for the GWs induced by the sub-Hubble modes, instead of the previous analytical approach used for the super-Hubble regime.

The power spectrum for the GWs induced by the sub-Hubble modes at time $\tau_*=-p_*^{-1}$ can be calculated by the formula \cite{Biagetti:2013kwa},
\begin{align} \label{Ph_Sub}
P^{\text{Sub}}_h(k,\tau_*)
&=
\frac{4}{\pi^4 M_p^4} k^3
\int^\infty_0 dp p^6 \int^\pi_0 d\theta \sin^5\theta 
\\& \times
\left|
\int^{\tau_*}_{\tau_i} d \tau_1 c_s^2(\tau_1) g_k(\tau_*,\tau_1)
\delta\phi_{p}(\tau_1) \delta\phi_{|\mathbf{k} - \mathbf{p}|}(\tau_1)
\right|^2 ~,\nonumber
\end{align}
where $ g_k(\tau_*,\tau_1) $ is the Green's function for the induced GWs, and $\delta\phi_{p}$ is the mode function of the quantum fluctuation $ \delta\hat{\phi}_\mathbf{p}$. The angle $ \theta $ is spanned by the wave vector $ \mathbf{k} $ and $\mathbf{p} $. Equation (\ref{Ph_Sub}) consists of two integrals, the phase space integral $\Delta\Pi=\int dp p^2 d\cos\theta d\varphi$ and the time integral. For the phase space integral, we use the thin-ring approximation (see Appendix \ref{ThinRing_APP}), while the integral over interacting time $\tau_1$ is performed in a numerical way.

In the SSR mechanism, the full form of the mode function $\delta\phi$ cannot be obtained analytically. Instead we use the subhorizon approximation and neglect the Hubble friction term in the EoM for the inflaton. After the approximation, the equation is of the standard Mathieu form and can be solved analytically in terms of Mathieu functions. The validity of the approximation depends on the physical wavelength of the considered mode. If its physical wavelength is small compared to the Hubble horizon, the approximation is expected to work well. Since here we are considering the sub-Hubble contributions, we only need to track $\delta\phi$ down to its horizon-crossing time. Henceforth, this approximation should be valid for the most region of integration.

To better organize the calculation, we introduce a dummy variable $s\equiv-p_* \tau_1$. Thus the time integral can be rewritten as
\begin{align} \label{Sub_Integral}
&\int^{\tau_*}_{\tau_i} d \tau_1 c_s^2(\tau_1)g_k(\tau_*,\tau_1)\delta \phi_{p}(\tau_1) \delta \phi_{|\mathbf{k} -\mathbf{p}|}(\tau_1)\\
\nonumber&=\frac{e^{iu+2iv}}{4p_*^5u^3}H^2\int_{1}^{v}ds \left[1-2\xi(1-\cos(2s))\right]M^2(s,v)\\
&\times e^{ius}\left[e^{-2iu}(1+iu)(-i-us)+e^{-2ius}(1-iu)(i-us)\right]~,\nonumber
\end{align}
where the function $M(s,v)$ is
\be
M(s,v)=\frac{S(s) \left(i C(v)-C'(v)\right)+C(s) \left(-i S(v)+S'(v)\right)}{-S(v) C'(v)+C(v) S'(v)}~.
\ee
As a shorthand notation, we denote the {Mathieu} sine and cosine {functions} as $C(x)\equiv C(1,\xi,x)$, $ S(x)\equiv S(1,\xi,x)$, and $C'(x)\equiv\partial C(1,\xi,x)/\partial x$, $S'(x)\equiv \partial S(1,\xi,x)/\partial x$.

Therefore the contribution from subhorizon modes is written as
\be \label{Ph_Sub_Form}
P^{\text{Sub}}_h(k,\tau_*)= \left\{
\begin{matrix}
	&16\xi^2\epsilon^2 A_s^2 \left(1-\frac{u^2}{4}\right)^2\frac{1}{u^4}|\mathcal{I}(u,v)|^2~,
	&u>\frac{\xi}{2},
	\\
	&32\xi \epsilon^2 A_s^2 \left(1-\frac{u^2}{4}\right)^2\frac{1}{u^3}|\mathcal{I}(u,v)|^2~,
	&u<\frac{\xi}{2},
\end{matrix}
\right.
\ee
where
\begin{align}
&\mathcal{I}(u,v) = \int_{1}^{v}ds \left[1-2\xi(1-\cos(2s))\right]e^{ius}\\
\nonumber&\times \big[ e^{-2iu}(1+iu)(-i-us)+e^{-2ius}(1-iu)(i-us) \big]\\
&\times \Big[ \frac{S(s) \left(i C(v)-C'(v)\right)+C(s) \left(-i S(v)+S'(v)\right)}{-S(v) C'(v)+C(v) S'(v)} \Big]^2~.\nonumber
\end{align}
Thus we arrive at the power spectrum $P^{\text{Sub}}_h(k,\tau_*)$ induced by the first major peak in the power spectrum $P_\zeta(k)$ \eqref{Pzeta}. The same treatments are applied to the analysis of the second and third resonance peaks in $P_\zeta(k)$ \eqref{Pzeta}. Note that the formula \eqref{Ph_Sub} and the thin-ring approximation can also be applied to the calculations of the power spectrum for the GWs induced by the super-Hubble modes at the end of inflation, by changing the expressions for the mode functions $\delta\phi_{p}$ and the interval of the integral to ($\tau_*$, $\tau_{\text{end}}$). These two approaches give the same results, which confirm our previous calculations in \eqref{Ph_PzetaPzeta_Inf} and \eqref{Finf}.

Note that the power spectrum for GWs induced by the sub-Hubble modes is controlled by $v=-p_*\tau_i$. This is the exponential of e-folding numbers between the triggering of resonance and the horizon exit of the characteristic inflaton mode. Larger $v$ gives a higher GW spectrum. However, this amplification cannot grow indefinitely. In order to keep the validity of our formalisms, we require $P_\zeta(p_*)<1$ and $P_h<1$. The former bound is considered in \cite{Cai:2018tuh} and gives
\be\label{PzetaCons}
A_s<\Big(\frac{p_*}{k_p}\Big)^{1-n_s} e^{-\xi v}~.
\ee
Notice that $p_*$ and $n_s$ do not explicitly enter the expression for $P^{\text{Sub}}_h(k,\tau_*)$ \eqref{Ph_Sub_Form}, but they do control the boundary of parameter region through \eqref{PzetaCons}.

\begin{figure}[h!]
	\centering
	\includegraphics[width=3.8in]{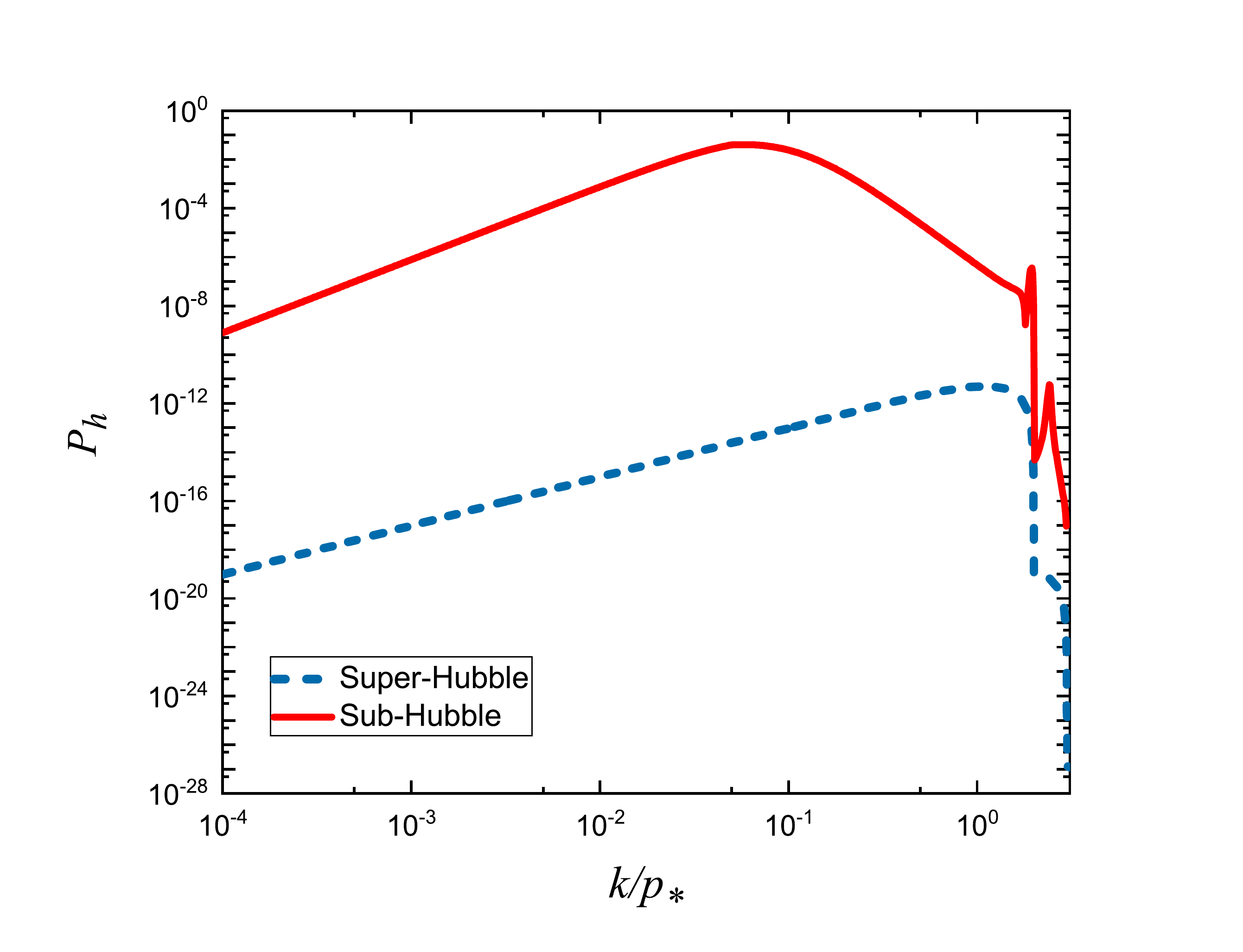}
	\caption{The power spectra for the GWs induced by super-Hubble modes (the blue dashed line) and sub-Hubble modes (the red line) of curvature perturbations with multiple peaks \eqref{Pzeta}. The GWs induced by the super-Hubble modes are estimated at the end of inflation $\tau_{\text{end}}$, while the GWs induced by the sub-Hubble modes are estimated at $\tau_*$, i.e., the Hubble crossing for the mode $p_*$. The average amplitude of the GWs induced by the sub-Hubble modes is about $10^{10}$ larger than that of the super-Hubble modes. The values of parameters we used as follows: $\epsilon = 10^{-3}$, $v = 200$, $ \xi = 0.1$ and $n_s=0.968$.}
	\label{Ph_Inflation_Fig}
\end{figure}

In Fig. \ref{Ph_Inflation_Fig}, we depict the power spectra in the super-Hubble regime and the sub-Hubble regime for a viable set of parameters indicated in the caption. We see that the average amplitude of the GWs induced by the sub-Hubble modes can be about $10^{10}$ larger than that of the super-Hubble modes. This gives us ample reasons to think that no significant corrections can arise from the evolution after $\tau_*$ (i.e. the super-Hubble regime). Physically speaking, it is also reasonable since the induced GWs possess a comparable wavelength as the characteristic mode. Therefore when the curvature perturbation of the characteristic mode $p_*$ is frozen outside the horizon, the induced GWs freeze as well. This conclusion is also confirmed by the numerical result, and thus the induced GWs from inflation is dominated by the contribution from the sub-Hubble regime, i.e., $P^{\text{Inf}}_h(k,\tau_{\text{end}}) \simeq P^{\text{Sub}}_h(k,\tau_*)$. 

Furthermore, the induced GWs from the inflationary era are large enough to be of observational interest and could be comparable to the induced GWs from the radiation-dominated era. We shall give more detailed discussion on this point in Sec. \ref{sec:OmegaGW}. Here, we notice that the peak of the induced GWs by the sub-Hubble modes is located around $\xi p_*/2$, which in our specific example takes $0.05 p_*$. Thus the location of the peak differs from the one of the induced GWs by the super-Hubble modes, as well as the one from the radiation-dominated era, both of which are located in the neighborhood of $p_*$. 
This shift of the peak is due to the interplay of phase integral and time integral in \eqref{Ph_Sub}. First, since the mode functions take the complicated forms in terms of Mathieu functions at the sub-Hubble scales; the integral \eqref{Sub_Integral} of mode functions would have significant effects on the location of the peak of the power spectrum $P^{\text{Sub}}_h(k,\tau_*)$. Secondly, at small $k$, the thin ring deforms to a shell due to the geometric effects; see Appendix \ref{ThinRing_APP} for details. Another different feature compared with the power spectrum induced by the super-Hubble modes is that the tail of power spectrum in the sub-Hubble regime has sharp peaks.

We comment that the large GWs induced by the sub-Hubble modes arise from two effects. First, the expansion of the Universe from the beginning of the sound speed oscillation to the Hubble crossing of {the mode $ p_* $}: $(\tau_i-\tau_*)/ \tau_* \simeq e^{\Delta N}$, where $\Delta N$ is the $e$-folding number for this period of inflation. The GWs induced by the super-Hubble modes are actually contributed from the integral $ \int^{\tau_{\text{end}}}_{\tau_*} $ in \eqref{Ph_Sub}, i.e., from the Hubble crossing of {the mode $ p_* $} to the end of inflation $\tau_{\text{end}}$: $ (\tau_*-\tau_{\text{end}}) / \tau_* \simeq 1$. In the SSR mechanism, $e^{\Delta N}\sim\mathcal{O}(100)$ is quite larger than 1. Another effect is during the late stage of the resonance, the mode function $\delta\phi_{p_*}(\tau)$ oscillates in a trigonometric way, giving rise to a nonvanishing central value of $\delta\phi_{p_*}^2$ which accumulates in the time integral in \eqref{Ph_Sub}.

\begin{figure}[h!]
	\centering
	\includegraphics[width=3in]{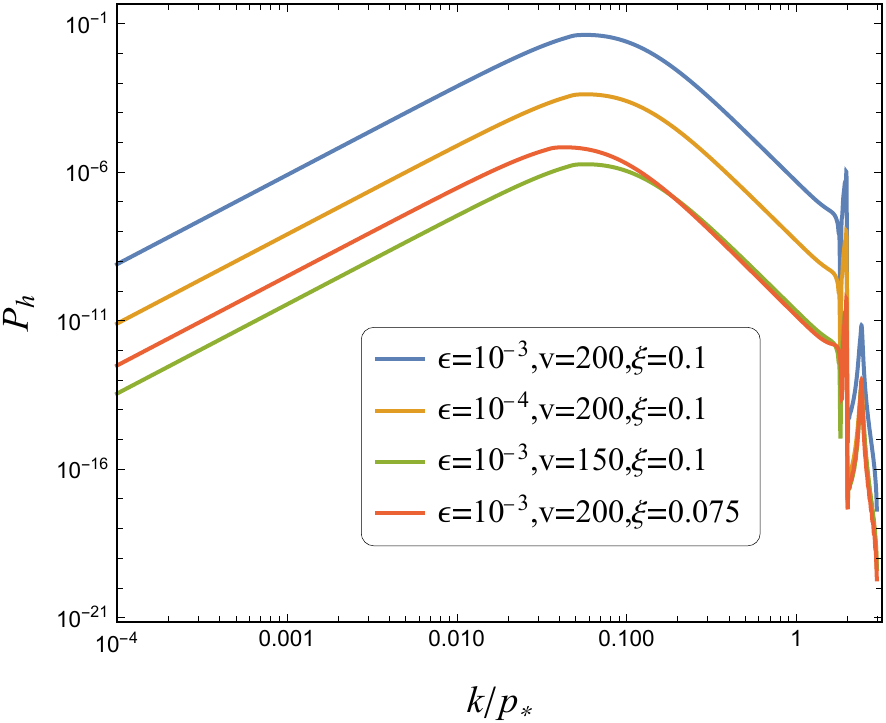}
	\caption{The inflationary power spectra of the induced GWs for different parameter choices. Since super-Hubble contributions are significantly smaller, here we can safely use sub-Hubble results to account for the whole $P^{\text{Inf}}_h(k,\tau_\text{end})$.}
	\label{PhComp_Fig}
\end{figure}

In Fig. \ref{PhComp_Fig}, we choose three sets of independent parameters $\epsilon$, $v$ and $\xi$ to study the dependence of inflationary power spectra $P^{\text{Inf}}_h(k,\tau_\text{end})$ on them. As expected, the power spectra are diminished by decreasing any of them while the shapes of the curves remain barely changed. This is reasonable since $v$ controls the duration of resonance and $\xi$ controls the rate of amplification. A decrease in either of them should cause $P^{\text{Inf}}_h(k,\tau_\text{end})$ to drop down. The slow-roll parameter $\epsilon$ appears in the overall normalization and has a simple quadratic dependence.

\section{Results} \label{sec:OmegaGW}

In the previous sections, we provided theoretical analyses of cosmological GWs induced by primordial curvature perturbations with multiple spikes \eqref{Pzeta} from the inflationary era to the radiation-dominated era. We derived the power spectra for the induced GWs, which are listed in Eqs. \eqref{Ph_PzetaPzeta_RD}, \eqref{Ph_PzetaPzeta_Inf} and \eqref{Ph_Sub}, respectively. Now we turn to forecast the observational implications on these induced GWs with the forthcoming GW experiments, e.g., the Laser Interferometer Space Antenna (LISA) \cite{Audley:2017drz}, Big-Bang Observer (BBO) \cite{Harry:2006fi}, Deci-hertz Interferometer Gravitational Wave Observer (DECIGO) \cite{Kawamura:2011zz}, and TianQin \cite{Luo:2015ght}. Moreover, the induced GWs at a low frequency band within the range of $[10^{-9}, 10^{-7}]{\rm Hz}$ may be accessible by the planned pulsar timing array experiments, e.g., the Square Kilometre Array (SKA) \cite{Janssen:2014dka} and International Pulsar Timing Array (IPTA) \cite{IPTA):2013lea}. To combine the observational windows of radio surveys and GW interferometers, the search for PBHs by virtue of probing the induced GWs is becoming promising in the era of multimessenger astronomy.

When the induced GWs associated with PBH formation evolve into the present along with the expansion of our Universe, they become a stochastic GW background which can be characterized by their energy spectra $ \Omega_{\text{GW}}(\tau_0,k) h_0^2 $ \cite{Maggiore:1999vm, Allen:1997ad}. Usually this is defined as the energy density of the GWs per unit logarithmic frequency, and $ h_0 $ is the reduced dimensionless Hubble parameter at the present time $ \tau_0 $. According to the definition of the effective energy for the GWs and the energy spectrum (see Appendix \ref{Basic_APP}), we get the relation between the energy spectrum and the power spectrum for the GWs [see \eqref{Omega_Ph_APP}] when the modes of GWs are well inside the Hubble horizon. For the induced GWs from the radiation-dominated era as shown in Eq. \eqref{Ph_PzetaPzeta_RD}, the energy spectrum observed today is estimated as \cite{Saito:2009jt,Bartolo:2018evs}
\be
\Omega_{\text{GW}}(\tau_0,k)
=
\Omega_{r,0} \Omega_{\text{GW}}(\tau_f,k) ~,
\ee
where $\tau_f$ is some time near the end of the radiation-dominated era and $\Omega_{r,0}$ is the present radiation energy density fraction. Since the energy density of GWs scales as radiation along with the cosmic expansion, the energy spectrum for the induced GWs does not dilute during the radiation-dominated era.

For the induced GWs during the inflationary era, which are given by Eqs. \eqref{Ph_PzetaPzeta_Inf} and \eqref{Ph_Sub}, the present energy spectrum can be approximately written as \cite{Boyle:2005se,Zhao:2006mm}
\be
\Omega_{\text{GW}}(\tau_0,k) \simeq 1.08 \times 10^{-6} P^{\text{Inf}}_h(k,\tau_{\text{end}}) ~,
\ee
for the frequency $ f > 10^{-10} $ Hz.

\begin{figure}
	\centering
	\includegraphics[width=7in]{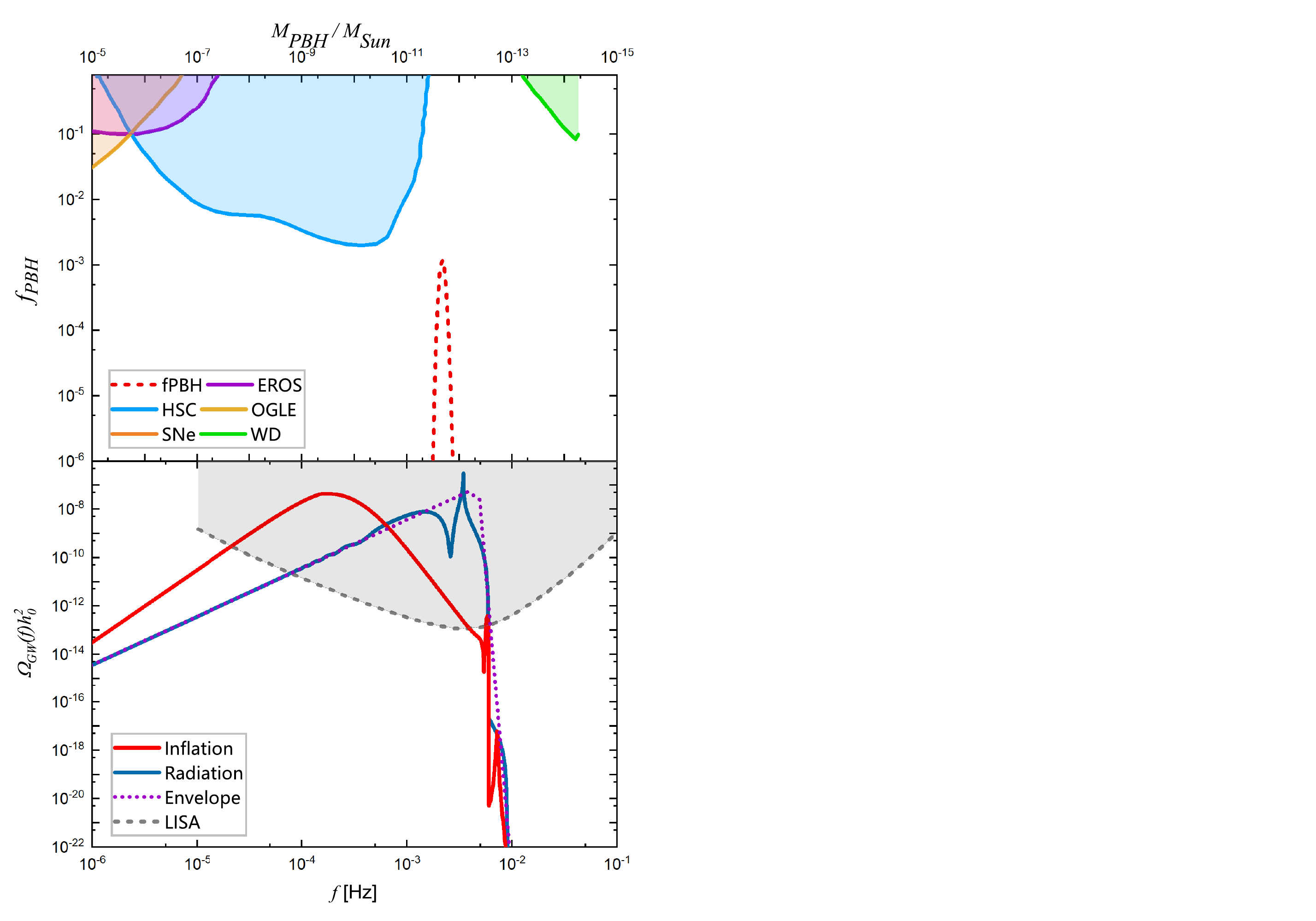}
	\caption{Upper panel: The $f_\mathrm{PBH}(M)$ spectrum (notice that the mass scale decreases from left to right) corresponding to the sensitivity of LISA (the grey region) \cite{Audley:2017drz}. Lower panel: The present energy spectra for the GWs with a limit by the sensitivity curve of LISA. The blue solid line denotes the spectrum of the induced GWs from the radiation-dominated era, and the purple dotted line is its envelope. The red solid line denotes the induced GWs from inflation, which is dominated by the contribution from the sub-Hubble regime as the previous discussions in Sec. \ref{sec:inducedGW}. Note that the present energy spectra of the induced GWs from inflation could be comparable to the induced GWs from the radiation-dominated era when we choose the values of parameters as follows: $\epsilon = 10^{-3}$, $v = 200$, $ \xi = 0.1$ and $n_s=0.968$. 
		Note that the frequency for the peak of $\Omega_{\text{GW}}(f)h^2_0$ from inflation is lower than the induced GWs from the radiation-dominated era. }
	\label{SpectraLISA_Fig}
\end{figure}

\begin{figure}
	\hspace*{-0.7cm}
	\includegraphics[width=7in]{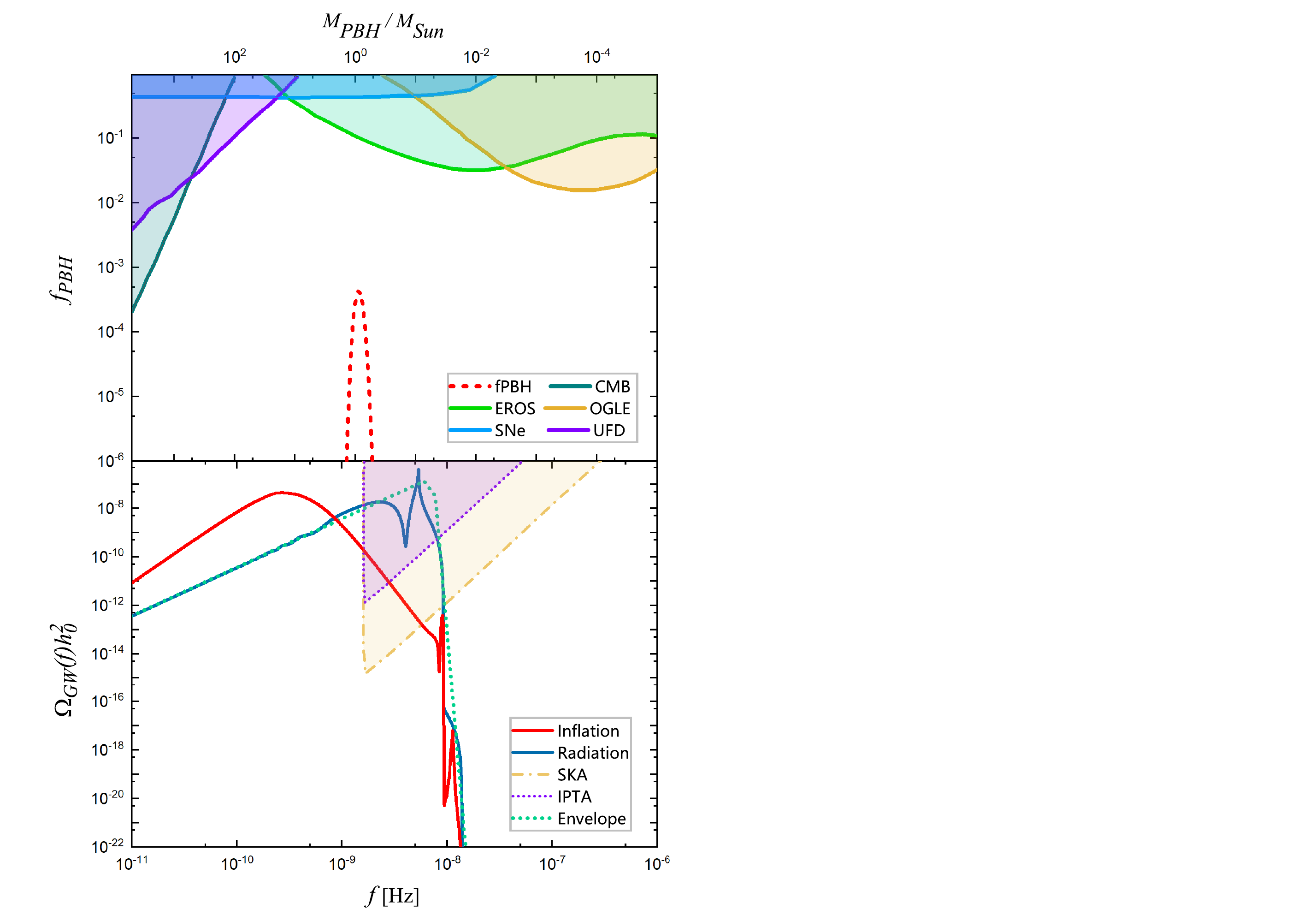}
	\caption{Upper panel: The $f_\mathrm{PBH}(M)$ spectrum (notice that the mass scale decreases from left to right) corresponding to the sensitivities of SKA (the yellow region) \cite{Santander-Vela:2018hjj} and IPTA (the purple region)\cite{IPTA):2013lea}. Lower panel: The present energy spectra for the GWs with a limit by the sensitivity curves of SKA and IPTA. The blue solid line denotes the spectrum of induced GWs from the radiation-dominated era, and the green dotted line is its envelope. The red solid line denotes the induced GWs from inflation. The values of the parameters we used are as follows: $\epsilon = 10^{-3}$, $v = 200$, $ \xi = 0.1$ and $n_s=0.968$.}
	\label{SpectraSKA_Fig}
\end{figure}

The present energy spectra $\Omega_{\text{GW}}(f)h_0^2$ of the induced GWs combined with the sensitivity curves of LISA, SKA and IPTA are shown in Fig. \ref{SpectraLISA_Fig} and Fig. \ref{SpectraSKA_Fig}, respectively. Also, the ratio $f_\mathrm{PBH}(M)$ of the energy density of PBHs to DM is another important phenomenological parameter. {In both figures, we also include the latest observational constraints on $f_\mathrm{PBH}$ from the corresponding electromagnetic windows \cite{Sato-Polito:2019hws}: i.e., constraints from white dwarf (WD) \cite{Graham:2015apa}; microlensing events (EROS/MACHO survey \cite{Tisserand:2006zx}, SubaruHSC \cite{Niikura:2017zjd}, OGLE \cite{Niikura:2019kqi}, and type Ia supernova SNe \cite{Zumalacarregui:2017qqd}); dynamic effects from ultra faint dwarf (UFD) \cite{Brandt:2016aco}; and CMB \cite{Ali-Haimoud:2016mbv, Poulin:2017bwe}.} Here the relation between PBH mass and the frequency of the induced GWs follows an inverse-square law, namely, $f_1^2 / f_2^2 \propto M_2 / M_1$. 
In order to show the correspondence between GW signals and PBH generation explicitly, we match the horizontal axes of GW frequency $f$ and PBH mass $M_{\rm PBH}$ in Fig. \ref{SpectraLISA_Fig} and Fig. \ref{SpectraSKA_Fig}.
Therefore $M_{\rm PBH}$ has a reversing axis direction.
For the sensitivity of LISA, we have chosen the scale $p_*\sim p_{\text{LISA}}\sim2\times10^{12}$Mpc$^{-1}$ associated a PBH of mass $10^{-12}M_{\text{Sun}}$ \cite{Bartolo:2018evs}, and the scale near the sensitivities of SKA and IPTA $p_*\sim 3 \times 10^{6} \text{Mpc}^{-1}$ corresponds to the PBHs with a mass of about $1M_{\text{Sun}}$.

In both Figs. \ref{SpectraLISA_Fig} and \ref{SpectraSKA_Fig}, we can see that the present energy spectra from the inflationary era are dominated by the sub-Hubble contribution, which are comparable to the present energy spectrum from the radiation-dominated era when we choose the values of parameters as follows: $\epsilon = 10^{-3}$, $v = 200$, $ \xi = 0.1$ and $n_s=0.968$. In Fig. \ref{SpectraLISA_Fig}, the energy spectra of the induced GWs, both from the inflationary era and the radiation-dominated era, exceed the sensitivity of LISA, and their peaks are located in the sensitive region of LISA. Moreover, the frequency of the peak of energy spectrum contributed from the inflationary era is around $10^{-1}$ lower than that of radiation-dominated era, so it is hopeful that we may distinguish these two signals in future LISA experiments. In Fig. \ref{SpectraSKA_Fig}, the induced GWs from the inflationary era and the radiation-dominated era both exceed the sensitivities of SKA and IPTA; nevertheless the signal from the inflationary era is smaller than the signal from the radiation-dominated era. 
However, the above conclusions are based on the choice of the characteristic scale $p_*$ in sound speed oscillation. If we choose an appropriate scale $p_*$,
the signal of the induced GWs from the inflationary era becomes detectable by LISA, SKA and IPTA. Hence, through the GW observations we can extract constraints on parameters $p_*$, $\xi$ and $v= -p_* \tau_i$ in the SSR mechanism.

The shapes of energy spectra of the stochastic GW background are crucial for the GW observations. 
As an approximation, we introduce the parametrization of the energy spectrum in the power-law form\footnote{The parametrization of the stochastic GW background is found to be powerful in probing the cosmic history, such as in \cite{Kuroyanagi:2018csn}.}:
\be
\Omega_{\text{GW}}(f) h_0^2
\sim
\left\{
\begin{matrix}
	A \left( \frac{f}{f_c} \right)^\alpha,~~~ f < f_c,
	\\
	A \left( \frac{f}{f_c} \right)^\beta,~~~ f > f_c ~,
\end{matrix} \right.
\ee
where $A$ represents the amplitude of energy spectrum at the critical frequency $f_c$, i.e., $\Omega_{\text{GW}}(f_c) h_0^2$. $\alpha$ and $\beta$ are the indices of the spectrum. For the induced GWs from the inflationary era, the parameters are given by
\begin{align}
&A \simeq 4.31 \times 10^{-7} , \nonumber
\\&
\alpha \simeq 3.0, ~~ \beta \simeq -5.4, ~~ f_c = 0.08 f_* ~.
\end{align}
For the induced GWs from the radiation-dominated era, we have
\begin{align}
&A \simeq 8.41 \times 10^{-8} ~(\text{LISA}), \nonumber
\\& A \simeq 2.05 \times 10^{-7} ~(\text{SKA}\&\text{IPTA}), \nonumber
\\&
\alpha \simeq 2.0, ~~ \beta \simeq -50.4, ~~ f_c = 1.6 f_* ~.
\end{align}
Note that because the term $(p_*/k_p)^{n_s-1}$ appears in the power spectrum \eqref{Ph_PzetaPzeta_RD}, the amplitudes of the parameterized energy spectra from the radiation-dominated era would change in different frequency ranges. As we have mentioned before, for the induced GWs from the sub-Hubble regime, the term  $(p_*/k_p)^{n_s-1}$ does not explicitly enter the expression for $P^\text{Sub}_h$ \eqref{Ph_Sub_Form}; hence the amplitudes of the parameterized energy spectra are the same in both frequency ranges of LISA and SKA (IPTA).


\section{Conclusion} \label{sec:conclusion}

In this article we perform a comprehensive analysis of the stochastic background of GWs induced nonlinearly by overly large primordial density perturbations that are accompanied with a process of PBH production. We report for the first time that the induced GWs can be resonantly enhanced within the sub-Hubble regime during inflation and hence make a significant contribution to the energy spectra that are of observable interest in the forthcoming GW experiments. Our study also confirms that the contribution of induced GWs in the super-Hubble regime during inflation is secondary due to the nature of slow-roll suppression. 
Accordingly, after summing up all contributions, the energy spectra of the induced GWs display a unique double-peak pattern that is innovative when compared with other works.
To develop the technique of computation, we proposed a novel parametrization of the power spectrum of primordial density perturbations with several spikes as inspired by the SSR mechanism. 
It is acknowledged that the first two peaks of the power spectrum would make the most important contribution to induce GWs nonlinearly. While the remaining peaks could continue to contribute at a secondary level, there is a steep tail on the profile of energy spectra at the high frequency band that damps out soon. This is manifestly different from the single peak case where the energy spectra were almost cut off at certain frequencies. In order to precisely characterize the energy spectra of the induced GWs from SSR mechanism, we put forward an envelope parametrization of their profiles which are expected to be measured in future astronomical observations.

Note that, depending on the mass scales of the PBHs, the characteristic peaks of energy spectra of the induced GWs may be located within the frequency band of $[10^{-5}, 10^{-2}]{\rm Hz}$, which is sensitive to the satellite projects of GW astronomy, or even as low as the frequency band of $[10^{-9}, 10^{-8}]{\rm Hz}$, which then is most relevant to radio astronomy. It is interesting to observe that, even for those PBHs with extremely light masses that have already been evaporated in the history of our Universe, they may leave a relic of induced GWs at a high frequency band, namely, within the LIGO range or even the kHz regime. This remarkable feature implies that the probe of the energy spectra of induced GWs in multimessenger astronomy has crucial implications for the search of PBHs at almost full frequency bands, even if those black holes may already disappear due to the mass loss via Hawking radiation.

Moreover, by comparing the observational abilities of GW astronomy with those of traditional telescopes upon the PBH mass spectrum, our results show explicitly that the observational window of GW instruments shall be more promising. Specifically, one could constrain the parameter space of the SSR mechanism by testing the energy spectrum of induced GWs within the scope of LISA, while the dominant peak of the corresponding mass spectrum might be far from the sensitivities of traditional telescopes. This fact, from another perspective, has well illustrated that the development of multimessenger astronomy shall become more and more promising in particular on the power of future GW detections.

In the end, we wish to highlight the implications of the present analysis on future studies from several perspectives. 
First of all, although our analysis is based on the SSR mechanism, from the perspective of methodology, the techniques developed in the present article can be easily extended into other scenarios with an extended mass spectrum for PBHs. 
Moreover our results indicate that, for the study of induced GWs, the traditional approach may be incomplete, since it only focused on the radiation dominated stage, but there can also be significant contributions from the sub-Hubble regime during inflation. 
Therefore it is worth checking the GWs signals from other PBH generation mechanisms, and also taking into account the contribution from the inflation stage.
Phenomenologically, an important lesson from our study is that the probe of the information about the very early Universe is no longer limited by the traditional CMB and LSS surveys, but also includes other astronomical telescopes at much smaller scales, as well as a brand new window of GW experiments. On the other hand, in the era of multimessenger astronomy, the search for PBHs are becoming more and more promising, making for a more and more serious DM candidate, which would inspire appropriate designs for the forthcoming experiments. In particular, a possible measurement of energy spectra of stochastic GW background with high precision could shed light on the nature of black holes existing in our Universe.


\

{\it \bf{Acknowledgments.}--}
We are grateful to A. Ach{\'u}carro, R. Brandenberger, B. Carr, C. Germani, S. Pi, M. Sasaki, and E. Saridakis
for stimulating discussions and valuable comments.
This work is supported in part by the National Youth Thousand Talents Program of China; by the NSFC ( Grants No. 11722327, No. 11653002, No. 11421303, No. J1310021 and No. 11961131007); by the CAST Young Elite Scientists Sponsorship (Grant No. 2016QNRC001); and by the Fundamental Research Funds for Central Universities.
D.G.W. is supported by a de Sitter Fellowship of the Netherlands Organization for Scientific Research (NWO).
D.G.W. is grateful to the hospitality of particle cosmology group at USTC where part of the work was carried out. 
All numerics are operated on the computer clusters LINDA \& JUDY in the particle cosmology group at USTC.\\

\appendix

\section{Basic Formulas for the Induced Gravitational Waves} \label{Basic_APP}

We review the basic formulas for the induced GWs \cite{Maggiore:1900zz, Boyle:2005se, Baumann:2007zm, Ananda:2006af}.
The EoM for the GWs $h_{ij}(\tau,\mathbf{x})$ induced by a source $\mathcal{S}_{lm}(\tau,\mathbf{x})$ can be written as
\be \label{EOM_hij_Sij_APP}
h''_{ij}(\tau,\mathbf{x}) + 2 \mathcal{H} h'_{ij}(\tau,\mathbf{x}) - \nabla^2 h_{ij}(\tau,\mathbf{x})
=
- 4 \hat{\mathcal{T}}^{lm}_{ij} S_{lm}(\tau,\mathbf{x}) ~,
\ee
where $\mathcal{H} \equiv a'/a$ is the comoving Hubble parameter, the prime denotes the derivative with respect to the conformal time $\tau$, and the projector operator $ \hat{\mathcal{T}}^{lm}_{ij} $ is used to extract the transverse and traceless part of the source, i.e.,
\begin{align} \label{TT_Operator_APP}
\hat{\mathcal{T}}^{lm}_{ij} \mathcal{S}_{lm}(\tau,\mathbf{x})
&= \sum_{\lambda = +,\times} \int \frac{\mathrm{d}^3 \mathbf{k}}{(2 \pi)^3} e^{i \mathbf{k} \cdot \mathbf{x}} \nonumber
\\& \times
e^\lambda_{ij}(\mathbf{k}) e^{\lambda}_{lm}(\mathbf{k}) \mathcal{S}_{lm}(\tau,\mathbf{k}) ~.
\end{align}
Note that repeated Latin indices refer to summation, and $\lambda = +, \times$ denote the polarizations of the GWs.
The polarization tensor $e^\lambda_{ij}(\mathbf{k})$ is expressed in terms of a pair of polarization vectors $e_i(\mathbf{k})$ and $\bar{e}_i(\mathbf{k})$, both of which are orthogonal to the wave vector $\mathbf{k}$:
\begin{align} \label{PolarizationTensor_APP}
e^+_{ij}(\mathbf{k}) =& \frac{1}{\sqrt{2}} \left[ e_i(\mathbf{k}) e_j(\mathbf{k}) - \bar{e}_i(\mathbf{k}) \bar{e}_j(\mathbf{k}) \right], \nonumber
\\
e^\times_{ij}(\mathbf{k}) =& \frac{1}{\sqrt{2}} \left[ e_i(\mathbf{k}) \bar{e}_j(\mathbf{k}) + \bar{e}_i(\mathbf{k}) e_j(\mathbf{k}) \right] ~.
\end{align}

Then one can write the EoM \eqref{EOM_hij_Sij_APP} for the GWs in Fourier space as
\be \label{EOM_hk_APP}
h^{\lambda''}_{\mathbf{k}}(\tau) + 2 \mathcal{H} h^{\lambda'}_{\mathbf{k}}(\tau) + k^2 h^\lambda_{\mathbf{k}}(\tau) = S^\lambda_{\mathbf{k}}(\tau) ~,
\ee
where
\be
S^\lambda_{\mathbf{k}}(\tau) = - 4 e^{\lambda}_{lm}(\mathbf{k}) \mathcal{S}_{lm}(\tau,\mathbf{k}) ~.
\ee
Here we take the Fourier transformations as
\be \label{S_Fourier_APP}
S_{lm}(\tau,\mathbf{k}) = \int \mathrm{d}^3 \mathbf{x} e^{- i \mathbf{k} \cdot \mathbf{x}} S_{lm}(\tau,\mathbf{x}) ~,
\ee
and
\be \label{hij_Fourier_APP}
h_{ij}(\tau,\mathbf{x}) = \sum_{\lambda = +,\times} \int \frac{\mathrm{d}^3 \mathbf{k}}{(2 \pi)^3}e^{i \mathbf{k} \cdot \mathbf{x}} h^\lambda_{\mathbf{k}}(\tau) e^\lambda_{ij}(\mathbf{k}) ~.
\ee

The particular solution of the EoM \eqref{EOM_hk_APP} is given by the Green function method,
\be \label{hk_Green_APP}
h^\lambda_{\mathbf{k}}(\tau)
=
\int^\infty_{-\infty} \mathrm{d} \tau_1 g_\mathbf{k}(\tau,\tau_1) S^\lambda_{\mathbf{k}}(\tau_1) ~,
\ee
where the Green's function $ g_\mathbf{k}(\tau,\tau_1) $ satisfies
\be
g''_{\mathbf{k}}(\tau,\tau_1) + 2 \mathcal{H} g'_{\mathbf{k}}(\tau,\tau_1) + k^2 g_{\mathbf{k}}(\tau,\tau_1) = \delta(\tau - \tau_1) ~.
\ee

The canonical quantization of the GWs \eqref{hij_Fourier_APP} can be written as $ \hat{h}^\lambda_{\mathbf{k}} = h^\lambda_k(\tau) \hat{a}^\lambda_\mathbf{k} + h^{\lambda*}_k(\tau) \hat{a}^{\lambda\dag}_{-\mathbf{k}} $, and the operator $ \hat{h}_{ij} $ is Hermitian, i.e., $  \hat{h}^\lambda_{\mathbf{k}} = \hat{h}^{\lambda\dag}_{-\mathbf{k}} $. The annihilation and creation operators $ \hat{a}^\lambda_\mathbf{k} $ and $ \hat{a}^{\lambda\dag}_\mathbf{k} $ satisfy the ordinary canonical commutation relations at the same time:
\begin{align}
&[\hat{a}^\lambda_\mathbf{k}, \hat{a}^{s\dag}_\mathbf{p}] = \delta^{\lambda s} (2 \pi)^3 \delta^{(3)}(\mathbf{k} - \mathbf{p}), \nonumber
\\&
[\hat{a}^\lambda_\mathbf{k}, \hat{a}^s_\mathbf{p}] = [\hat{a}^{\lambda\dag}_\mathbf{k},\hat{a}^{s\dag}_\mathbf{p}] =0 ~.
\end{align}
The correlator for the GWs is defined as
\be \label{Correlator_hk_APP}
\langle \hat{h}^\lambda_{\mathbf{k}}(\tau) \hat{h}^s_{\mathbf{k}'}(\tau) \rangle = (2 \pi)^3 \delta^{\lambda s} \delta^{(3)}(\mathbf{k} + \mathbf{k}') \frac{2 \pi^2}{k^3} P_h(k, \tau) ~,
\ee
where $ P_h(k, \tau) $ is the dimensionless power spectrum for the GWs of each polarization.

Similarly, the power spectrum for the comoving curvature perturbation $\zeta_{\mathbf{k}}(\tau)$ is given by
\be
\langle \hat{\zeta}_{\mathbf{k}}(\tau) \hat{\zeta}_{\mathbf{k}'}(\tau) \rangle = (2 \pi)^3 \delta^{(3)}(\mathbf{k} + \mathbf{k}') \frac{2 \pi^2}{k^3} P_\zeta(k, \tau) ~,
\ee

A stochastic background of the GWs can be characterised by its energy density fraction $ \Omega_{\text{GW}} $ \cite{Maggiore:1999vm,Allen:1997ad}, which is defined as the energy density of the GWs per unit logarithmic frequency,
\be \label{Omega_APP}
\Omega_{\text{GW}}(\tau,k) = \frac{1}{\rho_c(\tau)} \frac{\mathrm{d} \rho_{\text{GW}}(\tau,k)}{\mathrm{d} \ln k} ~,
\ee
where $ \rho_c(\tau) = 3 M_p^2 H^2(\tau) $ is the critical energy density at the conformal time $\tau$, and $H(\tau)$ is the Hubble parameter. The effective energy density of the GWs is usually defined as \cite{Boyle:2005se,Maggiore:1999vm}
\be
\rho_{\text{GW}} = \frac{1}{32 \pi G a^2(\tau)} \langle h'_{ij} h'_{ij} \rangle ~,
\ee
where the bracket means the time average over several periods of the GWs. 
When the relevant modes of the GWs are well inside the Hubble radius, one relates the $ \Omega_{\text{GW}}(\tau,k) $ and $P_h(k, \tau) $ as\cite{Boyle:2005se,Maggiore:1999vm}
\be \label{Omega_Ph_APP}
\Omega_{\text{GW}}(\tau,k) = \frac{1}{24} \left( \frac{k}{\mathcal{H}} \right)^2 \overline{P_h(k, \tau)} ~.
\ee
where the overbar denotes the time average over several periods of the GWs.

\section{Power Spectrum for the Induced Gravitational Waves} \label{PS_APP}

\subsection{Radiation-dominated era}\label{PS_RD_APP}

There are many early works on the induced GWs during the radiation-dominated era, and the details of calculations are presented in, e.g., \cite{Bartolo:2018rku,Baumann:2007zm,Ananda:2006af} and the references therein. Here we list the major formulas.

The relation between the first-order Bardeen potential and comoving curvature perturbations in the radiation-dominated era is
\be
\Phi_\mathbf{k}(\tau) \equiv \frac23 T(k\tau) \zeta_\mathbf{k} ~,
\ee
where the transfer function $ T(k\tau) $ can be solved from the EoM for the Bardeen potential
\be
T(z) = \frac{9}{z^2} \left[ \frac{\sin(z/\sqrt{3})}{z/\sqrt{3}} - \cos(z/\sqrt{3}) \right] ~,
\ee
where $ z = k\tau $.

We rewrite the source term (\ref{S_Fourier_APP}) as
\be \label{Sk_RD_APP}
S^\lambda_{\mathbf{k}}(\tau) = \frac{16}9 \int \frac{\mathrm{d}^3\mathbf{p}}{(2\pi)^3} \mathbf{e}^\lambda(\mathbf{k},\mathbf{p}) f(\mathbf{k},\mathbf{p},\tau) \zeta_\mathbf{p} \zeta_{\mathbf{k} - \mathbf{p}} ~,
\ee
where
\begin{align}
f(\mathbf{k},\mathbf{p},\tau)
&=
3 T(p\tau) T(|\mathbf{k} - \mathbf{p}|\tau)
+ \mathcal{H}^{-2} T'(p\tau) T'(|\mathbf{k} - \mathbf{p}|\tau) \nonumber
\\&
+ \mathcal{H}^{-1} T'(p\tau) T(|\mathbf{k} - \mathbf{p}|\tau) \nonumber
\\&
+ \mathcal{H}^{-1} T(p\tau) T'(|\mathbf{k} - \mathbf{p}|\tau) ~.
\end{align}
Note that the prime still denotes the derivative with respect to the conformal time $\tau$.

During the radiation-dominated era, it is more convenient to use the canonical form for $ h^\lambda_\mathbf{k}$, and the correlator for $h^\lambda_\mathbf{k}$ is expressed as
\begin{align} \label{Correlator_hk_RD_APP}
\langle \hat{h}^\lambda_\mathbf{k}(\tau) \hat{h}^s_{\mathbf{k}'}(\tau) \rangle
&=
\frac{1}{a^2(\tau)} \int^\tau \mathrm{d}\tau_1 \int^\tau \mathrm{d}\tau_2 g_k(\tau,\tau_1) g_k(\tau,\tau_2) \nonumber
\\& \times
a(\tau_1) a(\tau_2) \langle \hat{S}^\lambda_{\mathbf{k}}(\tau_1) \hat{S}^s_{\mathbf{k}'}(\tau_2) \rangle ~,
\end{align}
where the Green function satisfies the equation,
\be
g_k''(\tau,\tau_1) + \left( k^2 - \frac{a''(\tau)}{a(\tau)} \right) g_k(\tau,\tau_1) = \delta(\tau - \tau_1) ~,
\ee
and one can find the solution during the radiation-dominated era,
\be
g_k(\tau,\tau_1) = \frac{1}{k} \sin(k \tau - k \tau_1) \Theta(\tau - \tau_1) ~.
\ee

From the source term \eqref{Sk_RD_APP}, the two-point correlator $ \langle \hat{S}^\lambda_{\mathbf{k}}(\tau_1) \hat{S}^s_{\mathbf{k}'}(\tau_2) \rangle $ can be expressed by the four-point correlator $ \langle \hat{\zeta}_{\mathbf{p}} \hat{\zeta}_{\mathbf{k} - \mathbf{p}} \hat{\zeta}_{\mathbf{q}} \hat{\zeta}_{\mathbf{k}' - \mathbf{q}} \rangle $.  By the Wick's theorem, we obtain \cite{Baumann:2007zm,Ananda:2006af}
\begin{align}
\langle \hat{\zeta}_{\mathbf{p}} \hat{\zeta}_{\mathbf{k} - \mathbf{p}} \hat{\zeta}_{\mathbf{q}} \hat{\zeta}_{\mathbf{k}' - \mathbf{q}} \rangle
&=
\langle \hat{\zeta}_{\mathbf{p}} \hat{\zeta}_{\mathbf{q}} \rangle
\langle \hat{\zeta}_{\mathbf{k} - \mathbf{p}} \hat{\zeta}_{\mathbf{k}' - \mathbf{q}} \rangle \nonumber
\\&
+ \langle \hat{\zeta}_{\mathbf{p}} \hat{\zeta}_{\mathbf{k}' - \mathbf{q}} \rangle
\langle \hat{\zeta}_{\mathbf{k} - \mathbf{p}} \hat{\zeta}_{\mathbf{q}} \rangle ~.
\end{align}
Note that since $ \langle \hat{\zeta}_{\mathbf{p}} \hat{\zeta}_{\mathbf{k} - \mathbf{p}} \rangle = 0 $ with $ k \neq 0 $, this connection does not contribute to the four-point correlator $ \langle \hat{\zeta}_{\mathbf{p}} \hat{\zeta}_{\mathbf{k} - \mathbf{p}} \hat{\zeta}_{\mathbf{q}} \hat{\zeta}_{\mathbf{k}' - \mathbf{q}} \rangle $.

After having calculated the two-point correlator of the induced GWs \eqref{Correlator_hk_RD_APP}, we get the power spectrum for the induced GWs
\begin{align} \label{Ph_PzetaPzeta_RD_APP}
P^\text{RD}_h(k,\tau)
&=
\int^\infty_0 \mathrm{d}y \int^{1 + y}_{|1 - y|} \mathrm{d}x \left[ \frac{4 y^2 - (1 + y^2 - x^2)^2}{4 x y} \right]^2 \nonumber
\\& \times
P_\zeta(kx) P_\zeta ( k y ) F_{\text{RD}}(k,\tau,x,y) ~,
\end{align}
where
\begin{align}
&F_{\text{RD}}(k,\tau,x,y) \nonumber
\\
&=
\frac {4} {81} \frac{1}{z^2}
\big[ \cos^2 (z) \mathcal{I}_c^2 + \sin^2 (z) \mathcal{I}_s^2
+ \sin (2 z)  \mathcal{I}_c \mathcal{I}_s \big] ~.
\end{align}
and the functions $ \mathcal{I}_c $ and $ \mathcal{I}_s $ are given by
\begin{align} \label{Ic_APP}
\mathcal{I}_c( x , y ) &= 4 \int_{1}^{\infty} \mathrm{d} z_1  ( - z_1 \sin z_1 )
\Big\{ 2 T( x z_1 ) T( y z_1 ) \\ 
& + \big[ T( x z_1 ) + x z_1 T'( x z_1 ) \big] \big[ T( y z_1 ) + y z_1 T'( y z_1 ) \big] \Big\} ~, \nonumber \\
\label{Is_APP}
\mathcal{I}_s( x , y ) &= 4 \int_1^\infty \mathrm{d} z_1 ( z_1 \cos z_1 ) 
\Big\{ 2 T( x z_1 ) T( y z_1 ) \\
& + \big[ T( x z_1 ) + x z_1 T'( x z_1 ) \big] \big[ T( y z_1 ) + y z_1 T'( y z_1 ) \big] \Big\} ~, \nonumber
\end{align}
respectively, where $ z_1 = k \tau_1 $.

\subsection{Inflationary era}\label{PS_Infl_APP}

In the spatially flat gauge, one can write the source term in \eqref{EOM_hij_Sij_APP} contributed by the comoving curvature perturbation $\zeta$ as
\be \label{Slm_Infl_APP}
S_{lm}(\tau,\mathbf{x})
=
\frac{c_s^2(\tau)}{M_p^2} \partial_l \delta\phi(\tau,\mathbf{x}) \partial_m \delta\phi(\tau,\mathbf{x}),
\ee
where $ \delta\phi $ is the perturbed inflaton, and $ c_s $ is the sound speed of $ \delta\phi $. The $ \delta\phi $ relates to $ \zeta $ by $ \zeta = (H/\dot{\phi}_0) \delta\phi = (1/\sqrt{2 \epsilon} M_p) \delta\phi $. In the SSR mechanism, the modes of $ \delta\phi $ which are in the neighbourhood of the characteristic scale $ p_* $ can be exponentially enhanced by the narrow resonance effect, and they lead to the narrow peaks in the power spectrum $P_\zeta(k)$ \eqref{Pzeta}. The source term $S^\lambda_{\mathbf{k}}(\tau)$ in \eqref{EOM_hk_APP} during inflation reads
\be \label{Sk_Infl_APP}
S^\lambda_{\mathbf{k}}(\tau)
=
4 \frac{c_s^2(\tau)}{M_p^2}
\int \frac{\mathrm{d}^3 \mathbf{p}}{(2 \pi)^{3}} \mathbf{e}^\lambda(\mathbf{k},\mathbf{p})
\delta\phi_{\mathbf{p}}(\tau) \delta\phi_{\mathbf{k} - \mathbf{p}}(\tau).
\ee

It is sufficient to use the de Sitter (dS) approximation to get the Green function for the induced GWs during the inflationary era, so that $ a(\tau) = - 1/(H \tau) $ and $ \mathcal{H} = a'/a = -1/\tau $, and the Green function is given by \cite{Biagetti:2013kwa,Guzzetti:2016mkm}
\begin{align}
g_\mathbf{k}(\tau,\tau_1)
&=
\frac{1}{2 k^3 \tau_1^2} e^{-i k (\tau + \tau_1)}
\Big[e^{2 i k \tau} (1 - i k \tau) (- i + k \tau_1) \nonumber
\\&
+ e^{2 i k \tau_1} (1 + i k \tau) (i + k \tau_1) \Big] \Theta(\tau-\tau_1) ~.
\end{align}
The Green function only depends on the magnitude of wave vector $ \mathbf{k} $, i.e., $ k = |\mathbf{k}| $, and $ \Theta(\tau-\tau_1) $ is the Heaviside step function, such that $ \Theta(\tau - \tau_1) = 0 $ when $ \tau < \tau_1 $ and $ \Theta(\tau - \tau_1) = 1 $ when $ \tau > \tau_1 $. When we consider the modes of $ \delta\phi $ which are outside the Hubble horizon during inflation, the correlator for the induced gravitational waves is given by \eqref{hk_Green_APP},
\begin{align} \label{Correlator_hk_Infl_APP}
\langle \hat{h}^\lambda_{\mathbf{k}}(\tau) \hat{h}^s_{\mathbf{k}'}(\tau) \rangle
=
\int^\tau_{\tau_*} \mathrm{d}\tau_1 \int^\tau_{\tau_*} \mathrm{d}\tau_2
 ~ g_k(\tau,\tau_1) g_{k'}(\tau,\tau_2)
\langle \hat{S}^\lambda_\mathbf{k} \hat{S}^s_{\mathbf{k}'} \rangle ~,
\end{align}
where $ \tau_* = -1/p_* $ is the conformal time when the characteristic scale $ p_* $ exits the Hubble horizon during the inflationary era.

After calculations similar to the previous one \eqref{Ph_PzetaPzeta_RD_APP}, we acquire
\begin{align} \label{Ph_PzetaPzeta__Infl_APP}
P^\text{Super}_h(k,\tau)
&=
\int^\infty_0 \mathrm{d}y \int^{1 + y}_{|1 - y|} \mathrm{d}x \left[\frac{4 y^2 - (1 + y^2 - x^2)^2}{4 x y} \right]^2 \nonumber
\\& \times
P_\zeta(kx) P_\zeta ( k y ) F_{\text{Inf}}(k,\tau,u) ~.
\end{align}
Since the period of the sound speed oscillation $c_s$ is larger than the period from the Hubble crossing to the end of inflation, and the amplitude of oscillation (i.e., $\xi$) is also small, hence the function $F_{\text{Inf}}(k,\tau,u)$ is approximately
\begin{align}
F_{\text{Inf}}(k,\tau,u)
&\simeq
16 \epsilon^2 u^{-2}
\\& \times
\Big[ u + z \cos(z + u) - \sin(z + u) \Big]^2 ~, \nonumber
\end{align}
where $ u = k/p_* $. 
For the single-peak pattern of the power spectrum [i.e., considering the first major peak in \eqref{Pzeta}] $P_\zeta$, we get 
\begin{align} \label{Ph_Single_APP}
P^\text{Super}_h(k,\tau)
&\simeq
\frac14 \xi^2 \epsilon^2 A_s^2 e^{- 2 \xi p_* \tau_i} \left( \frac{p_*}{k_p} \right)^{2 n_s - 2} \nonumber
\\& \times
\Big[ u + z \cos(z + u) - \sin(z + u) \Big]^2 \nonumber
\\& \times
( 4 u^{-2} - 1 )^2 \Theta( 2 - u ) ~.
\end{align}

For the multi-spikes pattern of the power spectrum $P_\zeta$ \eqref{Pzeta}, we obtain
\begin{align} \label{Ph_Spikes_APP}
P^\text{Super}_h(k,\tau)
&\simeq \frac14 \xi^2 \epsilon^2 A_s^2 e^{- 2 \xi p_* \tau_i} \left( \frac{p_*}{k_p} \right)^{2 n_s - 2} 
\\& \times
\Big[ u + z \cos(z + u) - \sin(z + u) \Big]^2 P(u) ~, \nonumber
\end{align}
where the function $ P(u) $ is
\begin{align}
P(u)
&= ( 4 u^{-2} - 1 )^2 \Theta( 2 - u ) 
\\&
+ 4^{n_s - 3} a_2^2 ( 16 u^{-2} - 1 )^2 \Theta( 4 - u ) \nonumber
\\&
+ 9^{n_s - 3} a_3^2 ( 36 u^{-2} - 1 )^2 \Theta( 6 - u ) \nonumber
\\&
+ 2^{n_s - 3} a_2 \big[ 16 u^{-2} - ( 1 + 3 u^{-2} )^2 \big]^2 \Theta( 3 - u ) \Theta( u - 1 ) \nonumber
\\&
+ 3^{n_s - 3} a_3 \big[ 36 u^{-2} - ( 1 + 8 u^{-2} )^2 \big]^2 \Theta( 4 - u ) \Theta( u - 3 ) \nonumber
\\&
+ 2^{n_s - 3} a_2 \big[ 4 u^{-2} - ( 1 - 3 u^{-2} )^2 \big]^2 \Theta( 3 - u ) \Theta( u - 1 ) \nonumber
\\&
+ 3^{n_s - 3} a_3 \big[ 4 u^{-2} - ( 1 - 8 u^{-2} )^2 \big]^2 \Theta( 4 - u ) \Theta( u - 3 ) \nonumber
\\&
+ 6^{n_s - 3} a_2 a_3 \big[ 36 u^{-2} - ( 1 + 5 u^{-2} )^2 \big]^2 \Theta( 5 - u ) \Theta( u - 1 ) \nonumber
\\&
+ 6^{n_s - 3} a_2 a_3 \big[ 16 u^{-2} - ( 1 - 5 u^{-2} )^2 \big]^2 \Theta( 5 - u ) \Theta( u - 1 ) ~. \nonumber
\end{align}

We notice that the power spectra for the induced GWs in \eqref{Ph_Single_APP} and \eqref{Ph_Spikes_APP} exhibit the IR divergence. However, when the induced GWs evolve to the end of inflation, the power spectra are frozen and the IR divergence shifts to the much more lower frequency region, which does not affect the physical signals of the observable frequency ranges.

\subsection{The Thin-Ring Approximation}\label{ThinRing_APP}

\begin{figure}[h!]
	\centering
	\includegraphics[width=2.6in]{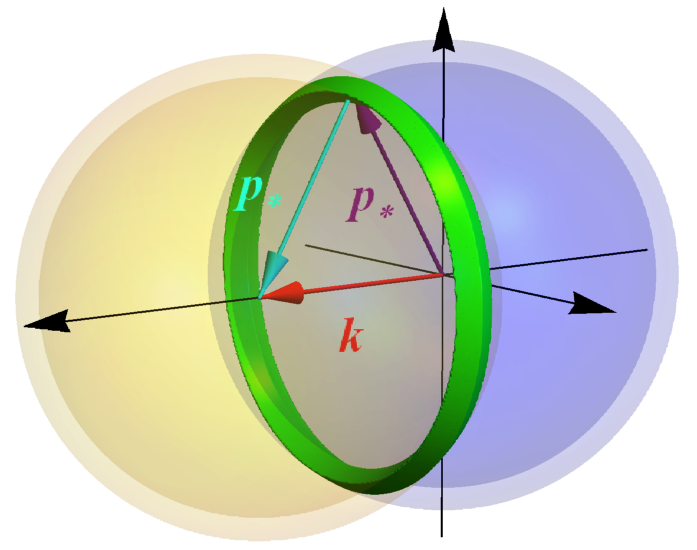}
	\caption{A sketch of the thin-ring approximation for the phase integral calculation. The phase integral is performed over the thin-ring region (in green).}
	\label{ThinFig_APP}
\end{figure}

In order to calculate the phase space integral $\Delta\Pi=\int dp p^2 d\cos\theta d\varphi $ in Eq. \eqref{Ph_Sub}, we use the thin-ring approximation. First, without loss of generality, we fix the direction of the wave vector $ \mathbf{k} $ of the induced GWs to lie along the z-direction. Then, considering the narrowness of the resonance band, and the fact that only those modes very close to it get amplified, we only need to integrate over a subspace in the entire $\mathbf{p}$ space. That is, the ringlike intersection of the two spheres centering at $\mathbf{0}$ and $\mathbf{k}$, each with a radius $p_*$. This region is shown in Fig. \ref{ThinFig_APP}. Notice that the cross section is not a round disk but a rhombic. Afterwards, simple geometrical calculation reveals the volume of the available phase space, namely, $\Delta\Pi= 2 \pi \xi^2 p_*^4 /k $. The approximation is based on the ringlike shape of the overlapping region. However, at small $k\lesssim \xi p_*$, geometric effects deform the ring into a shell and naturally cut off the IR divergence in $\Delta\Pi$, giving a finite result. Therefore, the full expression for $\Delta\Pi$ is written as $2 \pi \xi^2 p_*^4/k$ with $k>\frac{\xi}{2} p_*$ and $4 \pi \xi p_*^3$ with $k<\frac{\xi}{2} p_*$. Then the power spectrum \eqref{Ph_Sub} becomes
\begin{align}
P^{\text{Sub}}_h(k,\tau_*)
&=
\frac{4}{\pi^4 M_p^4} k^3
\times \frac{\Delta\Pi}{2\pi}p_*^4 \sin^4\theta
\\& \times
\left|
\int^{\tau_*}_{\tau_i} d \tau_1 c_s^2(\tau_1) g_k(\tau_*,\tau_1)
\delta\phi_{p}(\tau_1) \delta\phi_{|\mathbf{k} - \mathbf{p}|}(\tau_1)
\right|^2 ~,\nonumber
\end{align}
while the time integral over interacting time $\tau_1$ is performed numerically.

\bibliography{cosmo}

\end{document}